# Geometrical and Algebraic Approach to Central Molecular Chirality: a Chirality Index and an Aufbau Description of Tetrahedral Molecules

SALVATORE CAPOZZIELLO[1] AND ALESSANDRA LATTANZI[2]

[1]*Dipartimento di Scienze Fisiche and INFN (sez. di Napoli), Università di Napoli "Federico II", Complesso Universitario di Monte S. Angelo, Via Cinthia, I-80126, Napoli, Italy*

[2]*Dipartimento di Chimica, Università di Salerno, Via S. Allende, 84081, Baronissi, Salerno, Italy*

*ABSTRACT*    On the basis of empirical Fischer projections, we developed an algebraic approach to central molecular chirality of tetrahedral molecules. The elements of such an algebra are obtained from the 24 projections which a single chiral tetrahedron can generate in *S* and *R* absolute configurations. They constitute a matrix representation of the *O*(4) orthogonal group. According to this representation, given a molecule with *n* chiral centres, it is possible to define an "index of chirality $\chi \equiv \{n, p\}$", where *n* is the number of stereogenic centres of the molecule and *p* the number of permutations observed under rotations and superimpositions of the tetrahedral molecule to its mirror image. The chirality index, not only assigns the global chirality of a given tetrahedral chain, but indicates also a way to predict the same property for new compounds, which can be consistently built up.

Chirality is an underlying feature for physical systems ranging from elementary particles and molecules,[1] to macroscopic systems as spiral galaxies.[2] It can be discussed as a symmetry emerging in abstract spaces (such as the configuration space of particles) or in physical 3D-space of stereochemistry.[3] However, up today, an unifying view of this fundamental symmetry has not been achieved. The reason of this lack is, in our opinion, that a general theoretical approach has not been developed yet. In this report, we want to show that, at least for tetrahedral molecules, an algebraic description is realized considering the bonds as complex numbers in polar form and the empirical Fischer projections as the elements of a 4D configuration space. Such an approach leads to the definition of a chirality index capable of fully characterizing central molecular chirality of tetrahedral chains. Besides, this chirality index can be used to build up compounds in a sort of

Aufbau process where global chirality of products is consistently predicted. The paper is organized as follows. As first step, a tetrahedral molecule is geometrically described by complex numbers. The main feature which emerges from this description is the fact that the complex coniugation of two numbers can be directly related to the inversion between two bonds occurring in a chiral transformation. The final result of these considerations is that the chirality of a tetrahedral chain is assigned by the number $n$ of stereogenic centres and the number $p$ of permutations. Enantiomers, diastereoisomers and achiral molecules are easily classified through this rule. Moreover, we show that Fischer projections of a given tetrahedron are nothing else but the elements of the orthogonal $O(4)$ group which allows to represent rotations and inversions as matrix operators acting on the bonds.

Finally, taking into account all these results, an Aufbau rule can be deduced. In general, it allows to construct any simply connected tetrahedron assigning the chirality of the final compounds.

## GEOMETRICAL REPRESENTATION OF TETRAHEDRAL MOLECULES

The bonds of a tetrahedral molecule can be represented as complex numbers in polar form (1),

$$\Psi_j = \rho_j e^{i\theta_j} \qquad (1)$$

where $\rho_j$ is the length of the bond and $\theta_j$, the anomaly, is the relative position of the bond.[4]

The number $i = \sqrt{-1}$ is the imaginary unit. A tetrahedron is given by the sum vector of the four bonds.

$$\mathcal{M} = \sum_{j=1}^{4} \rho_j e^{i\theta_j} \qquad (2)$$

A tetrahedral molecule with a stereogenic centre can always be projected on a plane containing the stereogenic centre as in Fig. 1.

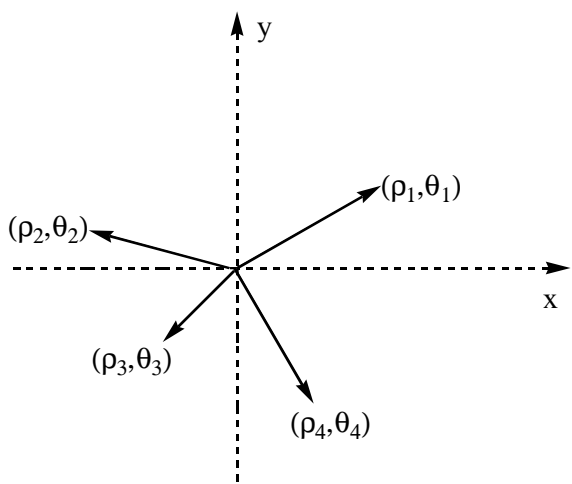

**Figure. 1.** Projection of a tetrahedral molecule on a plane containing the stereogenic centre.

If a molecule with one centre has four bonds, a molecule with two centres has seven bonds and so on. The general rule is

$$n = \text{centres} \quad \Leftrightarrow \quad 4n - (n-1) = 3n+1 \text{ bonds} \tag{3}$$

assuming simply connected tetrahedra. If atoms, acting as "spacers", are present between chiral centres, the number of bonds changes from $3n+1$ to $4n$, but the following considerations for consecutive connected tetrahedra remain valid. A molecule with $n$ stereogenic centres is then given by the sum vector

$$\mathcal{M}_n = \sum_{k=1}^{n} \sum_{j=1}^{3n+1} \rho_{jk} e^{i\theta_{jk}} \tag{4}$$

where k is the "centre-index" and j is the "bond-index". Again, for any k, a projective plane of symmetry is defined. The couple of numbers $\{\rho, \theta\} \equiv \{0,0\}$ assigns the centre in every plane.

Having in mind the definition of chirality, the behaviour of the molecule under rotation and superimposition has to be studied in order to check if the structure and its mirror image are superimposable. Chirality emerges when two molecules with identical structural formulas are not superimposable. Considering the geometrical representation reported in Fig.1, a possible situation is a rotation of 180° in the space around a generic axis passing through the stereogenic centre. Such an axis can coincide, for the sake of simplicity, with one of the bonds. After the rotation, two bonds result surimposable while the other two are inverted. The situation is depicted by the projection on the plane {x,y} as shown in Fig.2.

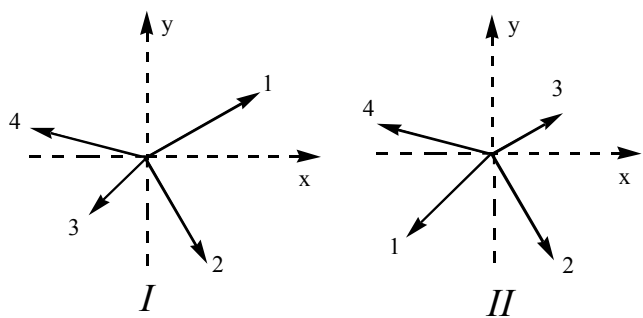

**Figure 2.** Picture of the projected situations before and after the superposition of a chiral tetrahedron over its mirror image. Groups 2 and 4 coincide while 3 and 1 are inverted. From a mathematical point of view, this is the result of the complex conjugation.

In formulae, for the inverted bonds, we have

$$\{\Psi_1 = \rho_1 e^{i\theta_1}, \Psi_3 = \rho_3 e^{i\theta_3}\} \Rightarrow \{\overline{\Psi}_1 = \rho_1 e^{i\theta_3}, \overline{\Psi}_3 = \rho_3 e^{i\theta_1}\} \quad (5)$$

In order to observe the reflection, the four groups must be of different nature. This simple observation shows that the chirality is related to the inversion of two bonds in the projective symmetry plane. Such a treatment can be repeated for any projective symmetry plane which can be defined for the $n$ centres. The possible results are that the molecule is fully invariant after rotation(s) and superimposition with respect to its mirror image (achiral); the molecule is partially invariant after rotation(s) and superimposition, i.e. some tetrahedra are superimposable while others are not (diastereoisomers); the molecule presents an inversion for each stereogenic centre (enantiomers).

The following rule can be derived: central chirality is assigned by the number $\chi$ given by the couple $n, p$ that is

$$\chi = \{n, p\} \quad (6)$$

where $\chi$ is the chirality index, $n$ is the principal chiral number and $p$ the secondary chiral number, that is $n$ is the number of stereogenic centres, $p$ is the number of permutations (at most one for any centre). The constraint

$$0 \leq p \leq n \quad (7)$$

has to hold.

This definition of chirality is related to the structure of the molecule and its properties under rotations and superimposition.

The sequence between achiral and chiral molecules is given by

$$\chi \equiv \{n, 0\} \quad \textit{achiral molecules}$$

$$\chi \equiv \{n, p < n\} \quad \textit{diastereoisomers}$$

$$\chi \equiv \{n, n\} \quad \textit{enantiomers}$$

As an example, the chirality of the amino alcohol in Fig. 3 is reduced to this rule. A permutation is observed for each stereogenic centre when testing superimposition of the structure **a** to **b**, hence $p = 2$, structure **a** is the enantiomer of **b** ( $\chi \equiv \{2, 2\}$ ).

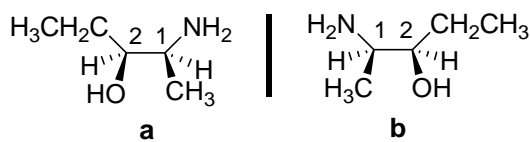

**Figure 3.** Structures of enantiomeric amino alcohols.

## ALGEBRAIC APPROACH DERIVED FROM FISCHER PROJECTIONS

Tetrahedral molecules can be represented by the planar projection formulas known as Fischer projections.[3] The ($S$)-lactic acid is reported in Fig. 4 as an example.

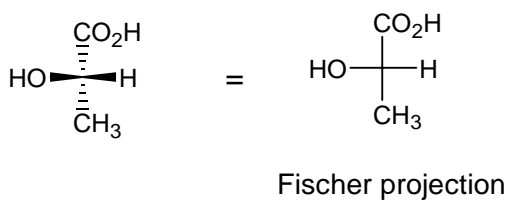

Fischer projection

**Figure 4.** Fischer projection of ($S$)-(+)-lactic acid.

By numbering from 1 to 4 the groups around the carbon centre as OH=1, $CO_2H$=2, H=3, $CH_3$=4, without taking into account the effective priorities of the groups,[3] there are 24 (=4! the number of permutations of 4 ligands among 4 sites) projections, twelve for the $S$-enantiomer (Fig. 5) and twelve for the $R$-enantiomer (Fig.6).

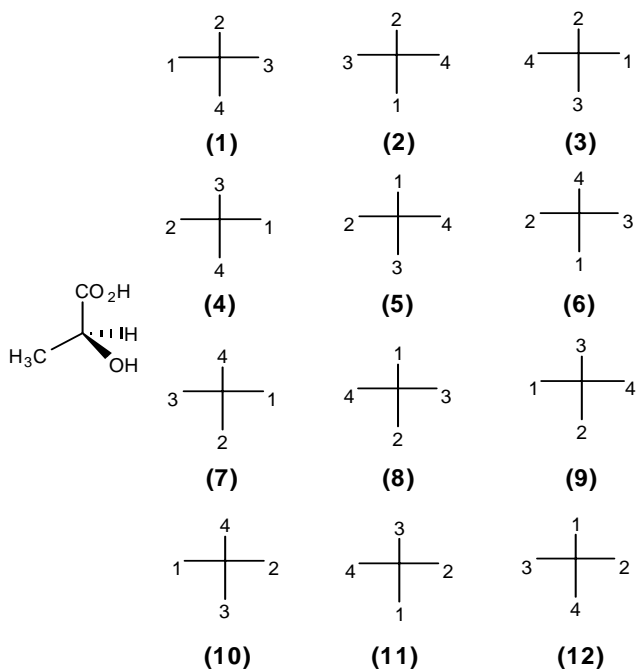

**Figure 5.** Twelve Fischer projections of (*S*)-(+)-lactic acid.

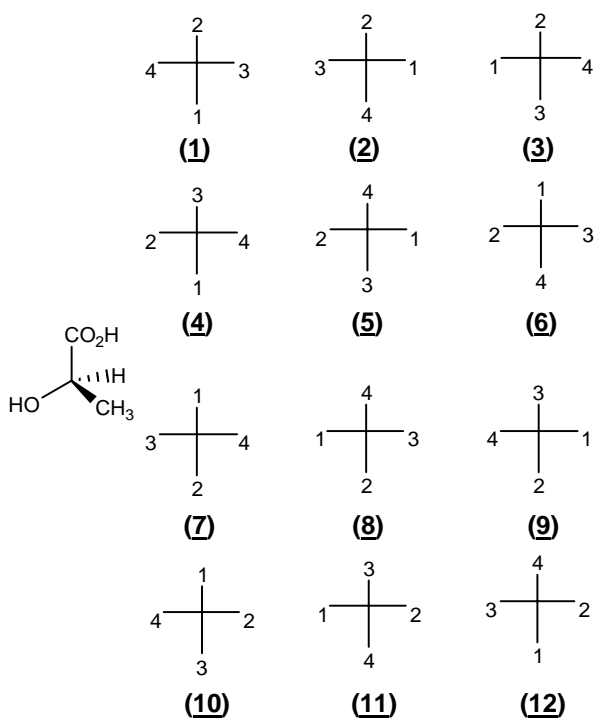

**Figure 6.** Twelve Fischer projections of (*R*)-(-)-lactic acid.

The permutations outlined in Fig. 6 derive from those in Fig. 5 simply by interchanging two groups. With these considerations in mind, it is immediate asking for an algebraic structure which can be built from Fischer projections. To reduce the Fischer rules to an algebraic structure, we define an operator $\chi_k$ acting on a tetrahedral molecule.[5] Take into account one tetrahedron, it can be assigned by a column vector $\mathcal{M}$, rewriting Eq. (2) as

$$\mathcal{M} = \begin{pmatrix} \Psi_1 \\ \Psi_2 \\ \Psi_3 \\ \Psi_4 \end{pmatrix} \tag{8}$$

$\Psi_j$ are defined in Eq. (1).

The corresponding Fischer projection is

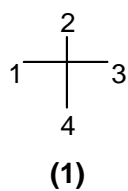

(1)

which is the first in Fig.5. The position of the bonds in the column vector (8) are assigned starting from the left and proceeding clockwise in the Fischer projection.

The matrix representation of the projection **(1)** is assumed as "fundamental", and the action of $\chi_1$ on the column vector $\mathcal{M}$ is the identity operator, that is

$$\chi_1 = \begin{pmatrix} 1 & 0 & 0 & 0 \\ 0 & 1 & 0 & 0 \\ 0 & 0 & 1 & 0 \\ 0 & 0 & 0 & 1 \end{pmatrix} \qquad \chi_1 \begin{pmatrix} \Psi_1 \\ \Psi_2 \\ \Psi_3 \\ \Psi_4 \end{pmatrix} = \begin{pmatrix} \Psi_1 \\ \Psi_2 \\ \Psi_3 \\ \Psi_4 \end{pmatrix} \tag{9}$$

The configuration **(2)** can be achieved by the operator $\chi_2$ acting as a rotation, that is

$$\chi_2 = \begin{pmatrix} 0 & 0 & 1 & 0 \\ 0 & 1 & 0 & 0 \\ 0 & 0 & 0 & 1 \\ 1 & 0 & 0 & 0 \end{pmatrix} \qquad \chi_2 \begin{pmatrix} \Psi_1 \\ \Psi_2 \\ \Psi_3 \\ \Psi_4 \end{pmatrix} = \begin{pmatrix} \Psi_3 \\ \Psi_2 \\ \Psi_4 \\ \Psi_1 \end{pmatrix} \tag{10}$$

By the same mechanism, it is possible to construct an operator $\overline{\chi}_k$ acting as an inversion, that is

$$\bar{\chi}_1 = \begin{pmatrix} 0 & 0 & 0 & 1 \\ 0 & 1 & 0 & 0 \\ 0 & 0 & 1 & 0 \\ 1 & 0 & 0 & 0 \end{pmatrix} \qquad \bar{\chi}_1 \begin{pmatrix} \Psi_1 \\ \Psi_2 \\ \Psi_3 \\ \Psi_4 \end{pmatrix} = \begin{pmatrix} \Psi_4 \\ \Psi_2 \\ \Psi_3 \\ \Psi_1 \end{pmatrix} \qquad (11)$$

By this approach, all the 24 projections can be obtained (12 for the (+) enantiomer and 12 for the (−) enantiomer):

Table I, (+)-enantiomer:

$$\chi_1 = \begin{pmatrix} 1 & 0 & 0 & 0 \\ 0 & 1 & 0 & 0 \\ 0 & 0 & 1 & 0 \\ 0 & 0 & 0 & 1 \end{pmatrix} \quad \chi_2 = \begin{pmatrix} 0 & 0 & 1 & 0 \\ 0 & 1 & 0 & 0 \\ 0 & 0 & 0 & 1 \\ 1 & 0 & 0 & 0 \end{pmatrix} \quad \chi_3 = \begin{pmatrix} 0 & 0 & 0 & 1 \\ 0 & 1 & 0 & 0 \\ 1 & 0 & 0 & 0 \\ 0 & 0 & 1 & 0 \end{pmatrix}$$

$$\chi_4 = \begin{pmatrix} 0 & 1 & 0 & 0 \\ 0 & 0 & 1 & 0 \\ 1 & 0 & 0 & 0 \\ 0 & 0 & 0 & 1 \end{pmatrix} \quad \chi_5 = \begin{pmatrix} 0 & 1 & 0 & 0 \\ 1 & 0 & 0 & 0 \\ 0 & 0 & 0 & 1 \\ 0 & 0 & 1 & 0 \end{pmatrix} \quad \chi_6 = \begin{pmatrix} 0 & 1 & 0 & 0 \\ 0 & 0 & 0 & 1 \\ 0 & 0 & 1 & 0 \\ 1 & 0 & 0 & 0 \end{pmatrix}$$

$$\chi_7 = \begin{pmatrix} 0 & 0 & 1 & 0 \\ 0 & 0 & 0 & 1 \\ 1 & 0 & 0 & 0 \\ 0 & 1 & 0 & 0 \end{pmatrix} \quad \chi_8 = \begin{pmatrix} 0 & 0 & 0 & 1 \\ 1 & 0 & 0 & 0 \\ 0 & 0 & 1 & 0 \\ 0 & 1 & 0 & 0 \end{pmatrix} \quad \chi_9 = \begin{pmatrix} 1 & 0 & 0 & 0 \\ 0 & 0 & 1 & 0 \\ 0 & 0 & 0 & 1 \\ 0 & 1 & 0 & 0 \end{pmatrix}$$

$$\chi_{10} = \begin{pmatrix} 1 & 0 & 0 & 0 \\ 0 & 0 & 0 & 1 \\ 0 & 1 & 0 & 0 \\ 0 & 0 & 1 & 0 \end{pmatrix} \quad \chi_{11} = \begin{pmatrix} 0 & 0 & 0 & 1 \\ 0 & 0 & 1 & 0 \\ 0 & 1 & 0 & 0 \\ 1 & 0 & 0 & 0 \end{pmatrix} \quad \chi_{12} = \begin{pmatrix} 0 & 0 & 1 & 0 \\ 1 & 0 & 0 & 0 \\ 0 & 1 & 0 & 0 \\ 0 & 0 & 0 & 1 \end{pmatrix}$$

Table II, (−) enantiomer:

$$\bar{\chi}_1 = \begin{pmatrix} 0 & 0 & 0 & 1 \\ 0 & 1 & 0 & 0 \\ 0 & 0 & 1 & 0 \\ 1 & 0 & 0 & 0 \end{pmatrix} \quad \bar{\chi}_2 = \begin{pmatrix} 0 & 0 & 1 & 0 \\ 0 & 1 & 0 & 0 \\ 1 & 0 & 0 & 0 \\ 0 & 0 & 0 & 1 \end{pmatrix} \quad \bar{\chi}_3 = \begin{pmatrix} 1 & 0 & 0 & 0 \\ 0 & 1 & 0 & 0 \\ 0 & 0 & 0 & 1 \\ 0 & 0 & 1 & 0 \end{pmatrix}$$

$$\overline{\chi}_4 = \begin{pmatrix} 0 & 1 & 0 & 0 \\ 0 & 0 & 1 & 0 \\ 0 & 0 & 0 & 1 \\ 1 & 0 & 0 & 0 \end{pmatrix} \quad \overline{\chi}_5 = \begin{pmatrix} 0 & 1 & 0 & 0 \\ 0 & 0 & 0 & 1 \\ 1 & 0 & 0 & 0 \\ 0 & 0 & 1 & 0 \end{pmatrix} \quad \overline{\chi}_6 = \begin{pmatrix} 0 & 1 & 0 & 0 \\ 1 & 0 & 0 & 0 \\ 0 & 0 & 1 & 0 \\ 0 & 0 & 0 & 1 \end{pmatrix}$$

$$\overline{\chi}_7 = \begin{pmatrix} 0 & 0 & 1 & 0 \\ 1 & 0 & 0 & 0 \\ 0 & 0 & 0 & 1 \\ 0 & 1 & 0 & 0 \end{pmatrix} \quad \overline{\chi}_8 = \begin{pmatrix} 1 & 0 & 0 & 0 \\ 0 & 0 & 0 & 1 \\ 0 & 0 & 1 & 0 \\ 0 & 1 & 0 & 0 \end{pmatrix} \quad \overline{\chi}_9 = \begin{pmatrix} 0 & 0 & 0 & 1 \\ 0 & 0 & 1 & 0 \\ 1 & 0 & 0 & 0 \\ 0 & 1 & 0 & 0 \end{pmatrix}$$

$$\overline{\chi}_{10} = \begin{pmatrix} 0 & 0 & 0 & 1 \\ 1 & 0 & 0 & 0 \\ 0 & 1 & 0 & 0 \\ 0 & 0 & 1 & 0 \end{pmatrix} \quad \overline{\chi}_{11} = \begin{pmatrix} 1 & 0 & 0 & 0 \\ 0 & 0 & 1 & 0 \\ 0 & 1 & 0 & 0 \\ 0 & 0 & 0 & 1 \end{pmatrix} \quad \overline{\chi}_{12} = \begin{pmatrix} 0 & 0 & 1 & 0 \\ 0 & 0 & 0 & 1 \\ 0 & 1 & 0 & 0 \\ 1 & 0 & 0 & 0 \end{pmatrix}$$

The matrices in Table I and II are the elements of a 4-parameter algebra.[6] Those in Table I are a representation of rotations, while those in Table II are inversions. Both sets constitute the group $O(4)$ of 4×4 orthogonal matrices. The matrices in Table I are the remarkable subgroup $SO(4)$ of 4×4 matrices with determinant +1. The matrices in Table II have determinant $-1$, being inversions (or reflections). They do not constitute a group since the product of any two of them has determinant +1. This fact means that the product of two inversions generates a rotation (this is obvious by inverting both the couples of bonds in a tetrahedron). The 24 matrices in Table I and II are not all independent. They can be grouped as different representations of the same operators. In fact all matrices, representing the same operator, have the same characteristic polynomial.[6] In other words, the characteristic equation of a matrix is invariant under vector base changes. In the (+) enantiomer case, the characteristic eigenvalue equation is

$$\det \| \chi_k - \lambda \mathbf{I} \| = 0 \tag{12}$$

where $\lambda$ is the eigenvalue and $\mathbf{I}$ is the identity matrix.

In the case of the (−) enantiomer, we have

$$\det\|\bar{\chi}_k - \lambda \mathbf{I}\| = 0 \qquad (13)$$

From Eqs. (12) and (13), we recover 6 eigenvalues

$$\lambda_{1,2} = \pm 1 \qquad \lambda_{3,4} = \pm i \qquad \lambda_{5,6} = \frac{-1 \pm i\sqrt{3}}{2} \qquad (14)$$

Inserting them into Eqs. (12) and (13), it is easy to determine the eigenvectors being

$$(\chi_k - \lambda \mathbf{I})\Psi = 0, \qquad (\bar{\chi}_k - \lambda \mathbf{I})\Psi = 0, \qquad (15)$$

with obvious calculations depending on the choice of $\chi_k$ and $\bar{\chi}_k$. It is worth noting that this approach, in particular Eqs. (12), (13) and the eigenvalues (14), can be related to the "pseudoscalar measurements" used to get information about the structure of chiral molecules in the sense indicated by Ruch.[7] In fact, a pseudoscalar, in particular its sign, can always be related to optical rotation or circular dichroism allowing the determination of the molecular chirality. The double sign of eigenvalues (14) is an indication in this sense, because it clearly points out the fact that a molecule is passing from a state with a given chirality to the opposite one. However, our $\lambda_s$ are not the Ruch's $\lambda_s$. In our case, $\lambda$ indicates the global chirality state of the molecule, in Ruch's case, it is a parameter describing the ligand. The stereochemical information, in our case, is relative to both enantiomers and it is not absolute.

These results can be extended to more general cases. For a molecule with $n$ stereogenic centres, we can define $n$ planes of projection and the bonds among the centres have to be taken into account. Eq. (4) can be written as

$$\mathcal{M}_n = \sum_{k=1}^{p} \overline{\mathcal{M}}_k + \sum_{k=p+1}^{n} \mathcal{M}_k \qquad (16)$$

where $\overline{\mathcal{M}}_k$ and $\mathcal{M}_k$ are generic tetrahedra on which are acting the operators $\bar{\chi}_l^k$ and $\chi_l^k$ respectively; k is the center index running from 1 to $n$; l is the operator index ranging from 1 to 12. For any tetrahedron, two possibilities are available:

$$\mathcal{M}_k = \chi_l^k \mathcal{M}_k^{(0)}, \qquad \overline{\mathcal{M}}_k = \bar{\chi}_l^k \mathcal{M}_k^{(0)} \qquad (17)$$

where $\mathcal{M}_k^{(0)}$ is the starting fundamental representation of the k-tetrahedron given by the column vector in Eq. (8). $\mathcal{M}_k$ and $\overline{\mathcal{M}}_k$ are the result of the application of one of the above matrix operators on the starting column vector $\mathcal{M}_k^{(0)}$. The index *p*, which, as previously stated, ranges $0 \leq p \leq n$, is the number of permutations which occur when the operators $\overline{\chi}_1^k$ act on tetrahedra. It corresponds to the number of reflections occuring in a *n*-centre tetrahedral chain. No inversions, but rotations occur when $\chi_1^k$ operators act on the molecule. Having this rule in mind, it follows that

$$\mathcal{M}_n = \sum_{k=1}^{n} \mathcal{M}_k, \qquad p = 0 \qquad (18)$$

is an achiral molecule;

$$\mathcal{M}_n = \sum_{k=1}^{p} \overline{\mathcal{M}}_k + \sum_{k=p+1}^{n} \mathcal{M}_k, \qquad 0 < p < n \qquad (19)$$

is a diastereoisomer since $[n-(p+1)]$ tetrahedra result superimposable after rotations, while *p*-ones are not superimposable, having, each of them, undergone an inversion of two of their bonds. Finally, an enantiomer results if

$$\mathcal{M}_n = \sum_{k=1}^{n} \overline{\mathcal{M}}_k, \qquad n = p \qquad (20)$$

where every tetrahedron results a mirror image of its starting situation after the application of any of the $\overline{\chi}_1^k$ operators. The chirality selection rule, geometrically deduced [Eq. (6)], is fully recovered. This selection rule enables a classification of tetrahedral chains by their chiral structure.

**MOLECULAR AUFBAU FOR TETRAHEDRAL CHAINS**

An *Aufbau* process can be derived.[8] The building-up process gives rise to a chirality index which assigns the intrinsic chiral structure of the final compound. As we have seen, the chirality index $\chi$ allows an immediate chiral characterization of a given tetrahedral chain. Let us take into account a molecule, which is well-defined in its chiral feature, in the sense that, considering also its mirror image, it is clear to assess if the molecule is an enantiomer, a diastereoisomer or an achiral molecule. After the addition of a further chiral centre to this structure and its mirror image, the resulting structure will be

$$\chi \equiv \{n+1, p+\Delta p\} \qquad (21)$$

where $\Delta p = 0,1$. The chiral properties of the new molecule are assigned by the $\Delta p$ value according to the following possibilities. If $\Delta p = 0$, we can have

$$\chi_s \equiv \{n,0\} \Rightarrow \chi_f \equiv \{n+1,0\} \qquad (22)$$

in this case, the starting compound is an achiral molecule as well as the final one.

Again, for $\Delta p = 0$, we can have

$$\chi_s \equiv \{n, p\} \Rightarrow \chi_f \equiv \{n+1, p\} \qquad (23)$$

in this case, the starting molecule is a diastereoisomer, being $n > p$, as well as the final structure.

Finally, if

$$\chi_s \equiv \{n, n\} \Rightarrow \chi_f \equiv \{n+1, n\} \qquad (24)$$

the starting molecule is an enantiomer, while the final one is a diastereoisomer.

If $\Delta p = 1$, the situations can be

$$\chi_s \equiv \{n,0\} \Rightarrow \chi_f \equiv \{n+1,1\} \qquad (25)$$

from an achiral molecule, a diastereoisomer is obtained;

$$\chi_s \equiv \{n, p\} \Rightarrow \chi_f \equiv \{n+1, p+1\} \qquad (26)$$

from a diastereoisomer, another diastereoisomer is obtained;

$$\chi_s \equiv \{n, n\} \Rightarrow \chi_f \equiv \{n+1, n+1\} \qquad (27)$$

from an enantiomer, we get another enantiomer.

Eqs. (22)-(27) take into account all the possibilities, which can be easily iterated adding up any number of chiral centres to a given chain. In the general case, the *Aufbau rule* is

$$\chi \equiv \{n+n', p+p'\}; \forall n' \geq 1, \quad p' = \sum_{j=1}^{n'} \Delta p_j, \quad \Delta p_j = 0,1 \qquad (28)$$

However, we have to consider that the rule works only for simply connected tetrahedral chains, where the chiral features are well-establishd with respect to the mirror image. In this sense, chirality is not an absolute feature of the molecules. Adding up a chiral centre to a structure gives rise to a new molecule, where $\chi \equiv \{n+1, p+\Delta p\}$. The fact that, in the addition, the variation of $p$ can be $\Delta p = 0,1$ assigns the chiral feature of the new compound.

An example of the building-up process is reported in Fig.7.

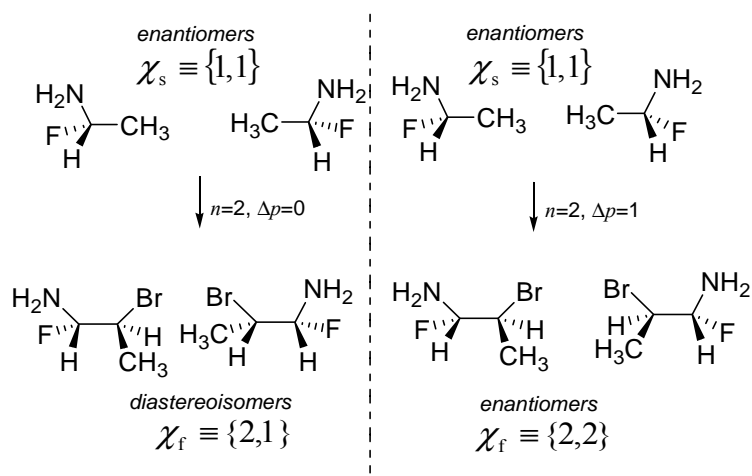

**Figure 7.** Aufbau process consisting in adding up a chiral centre to a given chiral tetrahedron.

In Fig.8, the degenerate example for the *meso*-tartaric acid is reported, where two chiral centres are identical, introducing a further degree of symmetry to the final structure. In this case, the situation $\chi_f \equiv \{2,2\}$ for $\Delta p = 1$ is equivalent to $\chi_f \equiv \{2,0\}$, since the two molecules are superimposable, hence the structure is achiral.

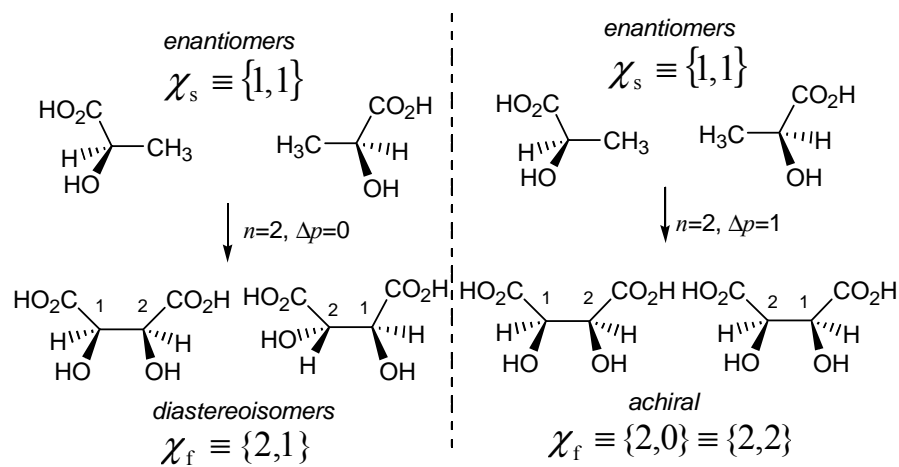

**Figure 8.** A degenerate example of Aufbau process consisting in adding up a chiral centre, having identical substituents of the starting chiral tetrahedron.

Last consideration indicates that such an Aufbau approach is working only if the chiral centres are different and, in this respect, the procedure is suitable for the description of chiral structures.

**CONCLUSION**

Central molecular chirality can be dealt under the standard of an algebraic approach where any configuration of tetrahedral chains can be achieved by the matrix operators of $O(4)$ group. Geometrically, this means that molecular rotations and inversions are realized by representiong the molecule as the superposition of complex numbers in polar form, while the chiral structure is given by the number of stereogenic centres and the number of permutations of couples of bonds (both these numbers constitute the "chiral index").

As a final remark, it is worth stressing that the outlined approach could constitute a way to achieve an unifying view of chirality since it is possible to show that it works for microscopic structures as elementary particles and huge macroscopic systems as spiral galaxies.